\documentclass[aps,pra,twocolumn,groupedaddress,showpacs,showkeys,superscriptaddress]{revtex4-1}

\usepackage{amsmath}
\usepackage{amssymb}
\usepackage{graphicx}
\usepackage{epstopdf}

\usepackage{subcaption}
\usepackage{mwe}

\usepackage{media9}
\usepackage{hyperref}

\begin{document}

\title{High-quality activation function for applications in neuromorphic photonic chips realized using low-quality nonlinear optical resonators}

\author{Ivan A. Pshenichnyuk}
\email[correspondence address: ]{i.pshenichnyuk@skoltech.ru}
\author{Kamil R. Taziev}
\author{Sergey S. Kosolobov}
\author{Vladimir P. Drachev}
\affiliation{Skolkovo Institute of Science and Technology, Moscow 121205, Russian Federation}

\date{\today}

\begin{abstract}
Integrated optical devices that can realize a threshold filtration of signals are in demand in photonics. In particular, they play a key role in neuromorphic chips, acting as optical neurons. A list of requirements exist for thresholders to be practically applicable in this context. A value of the threshold, in general, should be independently tunable for each neuron. A sharpness of the corresponding step function should be also dynamically variable, to allow switching between deterministic and stochastic algorithms, for example, in optical Ising machines. Nonlinear ring resonators draw the attention of researchers in this field since they potentially can provide the required type of threshold behavior. Here we suggest the switching mechanism that implements the property of resonators to provide extremely sharp (with respect to the wavelength) $\pi$ phase shifts near the critical coupling regime. Adding a variable source of losses into the resonator, a well controlled broadening of the threshold can be achieved. Sharpness of the step in this case is independent on the quality factor of resonators and corresponding width of resonances. The nonlinear switching mechanism allows to use this feature to construct efficient optical neurons, that operate at small intensities. It also allows to use conventional materials like silicon with a relatively weak Kerr nonlinearity. The obtained results potentially lead the way to fast nonlinear Kerr effect based activation functions that can operate in continuous wave regime.
\end{abstract}

\maketitle


Optical microresonators play a significant role in integrated photonics. Silicon waveguides evanescently coupled to ring or disk resonators suggest numerous applications in photonic circuits \cite{bogaerts-2012}. One of the most obvious examples is a resonator based filter. At certain resonant wavelengths, destructive interference between multiple waves, passing various ways in a resonator cavity, cause the absence of transmittance at the output of the coupled waveguide, acting similar to Fabry–Perot etalon \cite{yariv}. 
Filtering effects become particularly interesting when a nonlinearity in the ring is involved. The dependence of transmittance on intensity of a passing wave in a vicinity of a resonance may be quite peculiar \cite{rukhlenko-2010, chen-2012}. Nontrivial shapes can be used to create photonic nonlinear elements, like thresholders and logic gates \cite{huang-2019,xu-2007}. It is also suggested to use it in neuromorphic photonic chips \cite{tezak-2020,tezak-2015,huang-2021}.

Particularly important for applications in the field of photonic artificial intelligence are elements that can realize the activation function (or logistic function) \cite{hertz, prucnal, ertel}. They are used to check if a signal collected by a neuron is above or below a certain threshold. There exist a list of requirements for such an element to make them practical. In order to allow neuromorphic algorithms, like for example Ising machines \cite{kielpinski-2016,newman}, to solve a general class of problems, a value of a threshold must be dynamically tunable. It becomes especially important in optimization problems, where each photonic neuron should have its own threshold. Moreover, there must be a possibility to switch between sharp and smooth step-functions, with a well controlled degree of smearing. It is used to switch between deterministic and stochastic regimes and vary a degree of chaos in the system. In a mathematical sense, the perfect shape for an activation function is provided by the Fermi-Dirac distribution with variable "chemical potential" and "temperature" \cite{hertz}. From this point of view the curves provided by intensity dependent transmittance of nonlinear resonators are not sufficiently good \cite{tezak-2020}.

In search of an appropriate optical signal transformation mechanism it was discovered that the phase variation in resonator based filters deserve a particular attention. The phase shift is quite strong in a vicinity of a resonance. Thus, a nonlinear variation of a resonant condition may cause an intensity dependent switching that generates the required activation curve with desired properties \cite{campo-2022}. The phase switching can be easily transformed into an amplitude switching using interference effects. One of the ways is the integration of a ring resonator into an arm of a Mach-Zehnder interferometer (MZI) \cite{huang-2021,tezak-2020}.

An important characteristics of resonators is a width of resonances, that is directly linked with their quality factors and losses. In general, a large width makes the nonlinear switching mechanism problematic. It requires to apply a significant intensity to modify the resonant condition strongly enough to provide shifting from one side of the resonance to another. This statement is valid for both amplitude and phase switching. Large switching intensity is impractical from the point of view of industrialization and stability of devices. To decrease the switching intensity, materials with strong nonlinear properties can be used \cite{pshenichnyuk-2024}. Unfortunately, silicon is not one of them. Increasing the quality of resonators one may decrease the impact of both above mentioned factors, but the production of such resonators can be challenging.

The phase shifting behavior of ring resonators becomes special close to the critical coupling regime \cite{heebner-2004, liu-2008, chalupnik-2023}. Instead of a $2\pi$ phase shift smeared over the resonance, one gets sharp $\pi$ shift in the middle of the resonance. Exactly in the critical coupling regime we expect to obtain a perfect step function, that can be broadened using small deviations from the regime. This broadening is quite thin and it has nothing to do with the thickness of the resonances and their quality factor. Small thickness allows to switch the system using weak nonlinear materials (like silicon) at relatively small intensities. In this letter we explore the possibility to use this effect for the realization of activation functions in neuromorphic photonic chips \cite{shen-2017, harris-2018}.


A scattering matrix $S$ of a ring resonator based filter with one coupled waveguide can be expressed in the following form \cite{yariv}
\begin{equation}
  S = \frac{ae^{i\varphi}+ib}{1-iabe^{i\varphi}}.
  \label{rr_linear_smatrix}
\end{equation}
It links input and output complex wave amplitudes as $A_{out}=SA_{in}$. Here $\varphi=2\pi Ln_{eff}/\lambda$ is a phase obtained after one travel around the ring, $L$ - is the length of the resonator, $n_{eff}$ - effective index of a waveguide mode, $\lambda$ - wavelength. A parameter $a$ varies between $0$ and $1$ and defines losses in the ring ($1$ corresponds to a lossless ring), while $b=\sqrt{\eta}$ defines a coupling strength between the waveguide and the ring. A splitting ratio $\eta$ is defined as a fraction of power that goes straight through the coupler, thus $(1-\eta)$ is a fraction coupled to the ring. Computing an absolute square of Eq.~\ref{rr_linear_smatrix} one obtains the well known formula for a transmittance of a ring resonator based filter \cite{yariv}. Correspondingly, a complex argument of Eq.~\ref{rr_linear_smatrix} yields the phase shift. The critical coupling condition in our notations is written as $a=b$. It is easy to check that the in-resonance transmittance goes strictly to zero in this case.

\begin{figure}
\centerline{\includegraphics[width=0.5\textwidth]{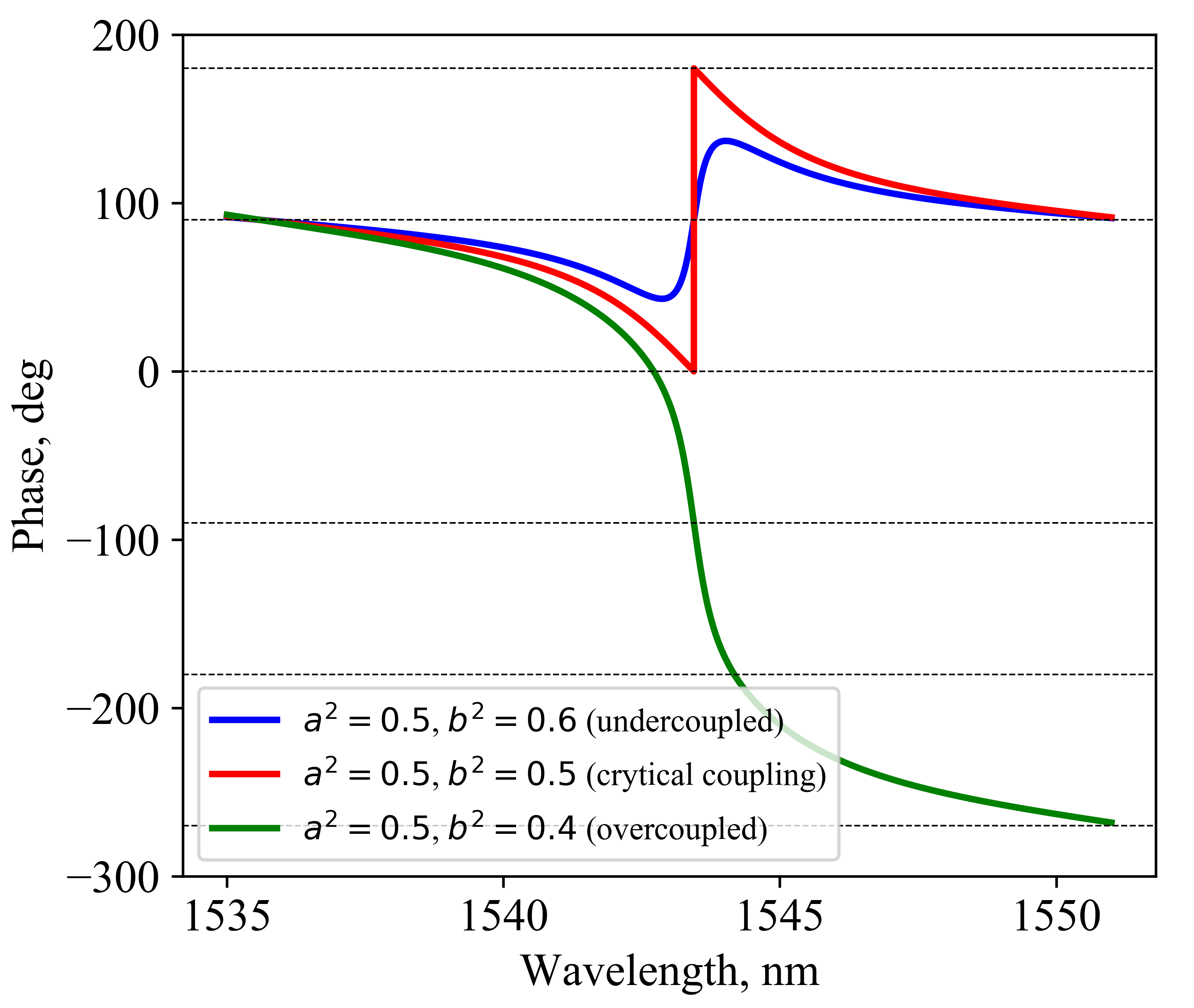}}
\caption {Three types of phase variation occurring in ring based filters near a resonant wavelength.
\label{fig1_rrlinphases}}
\end{figure}

In Fig.~\ref{fig1_rrlinphases} three types of a phase behavior near a resonance are demonstrated. Exactly in the critical coupling regime (red curve), when losses are balanced with coupling, there is one sharp $\pi$ phase shift in the center of the resonance. It can be made smoother in undercoupled regime, when $b>a$ (blue curve). Quite diferently, in overcoupled regime, when $a>b$, we obtain smooth $2\pi$ phase shift (green curve). In the latter case the variation of phase takes place in the range of wavelengths that coincides with the width of the resonance. In other two cases the thickness of sharp $\pi$ phase shift is defined by the proximity to the critical coupling regime and does not depend on the width of the resonance and its quality factor. 

In general, a quality factor of a resonator is expressed as a frequency of a resonance divided by its full width at half maximum (expressed in Hz). In our analytical model it is predefined by parameters $a$ and $b$.  For instance, for $a^2=b^2=0.5$ the quality factor is $472$, which is rather low, characterized by broad resonances in the spectrum. Taking another set of parameters like, for example, $a^2=b^2=0.9$ (used in Fig.~\ref{fig3_rrlin_cc}) we can slightly increase the quality up to $2943$, but it is still considered as low, compared with modern experimental achievements \cite{burla-2015}.

In a nonlinear resonator the refractive index of its material depends on the signal amplitude, which we denote as $A_r$. This fact results in additional amplitude dependent phase shift in the ring. It can be shown that for the minimalistic practical theory of nonlinear resonators it is sufficient to assume a quadratic dependence
\begin{equation}
  \varphi(|A_r|^2) = \frac{2\pi L}{\lambda}n_{eff} + \gamma |A_r|^2,
  \label{rr_nonlin_phase}
\end{equation}
with a nonlinear coefficient $\gamma$. If one explicitly refers to Kerr nonlinearity in the ring, coefficient $\gamma$ can be linked with corresponding $n_2$ coefficient by specifying the mode polarization and the waveguide structure.

It becomes challenging to obtain an analytical expression for the $S$-matrix in the nonlinear case, since the phase depends on $A_r$. Nevertheless, it is possible to write the solution in the parametric form 
\begin{equation}
  A_{in} = A_r \frac{1-iabe^{i\varphi(|A_r|^2)}}{\sqrt{1-b^2}},
  \label{rr_nonlin_smatrix1}
\end{equation}
\begin{equation}
  A_{out} = A_r \frac{ae^{i\varphi(|A_r|^2)}+ib}{\sqrt{1-b^2}}.
  \label{rr_nonlin_smatrix2}
\end{equation}
Allowing $A_r$ to pass a certain range of values one obtains numerically both $A_{in}$ and $A_{out}$ and thus can reconstruct the $S$-matrix. Since $A_r$ in Eqs.~\ref{rr_nonlin_smatrix1}-\ref{rr_nonlin_smatrix2} stands under the argument of a periodic function, in certain cases multiple combinations exist that correspond to bistable regimes of the resonator \cite{rukhlenko-2010, haus, boyd}.

\begin{figure}
\centerline{\includegraphics[width=0.5\textwidth]{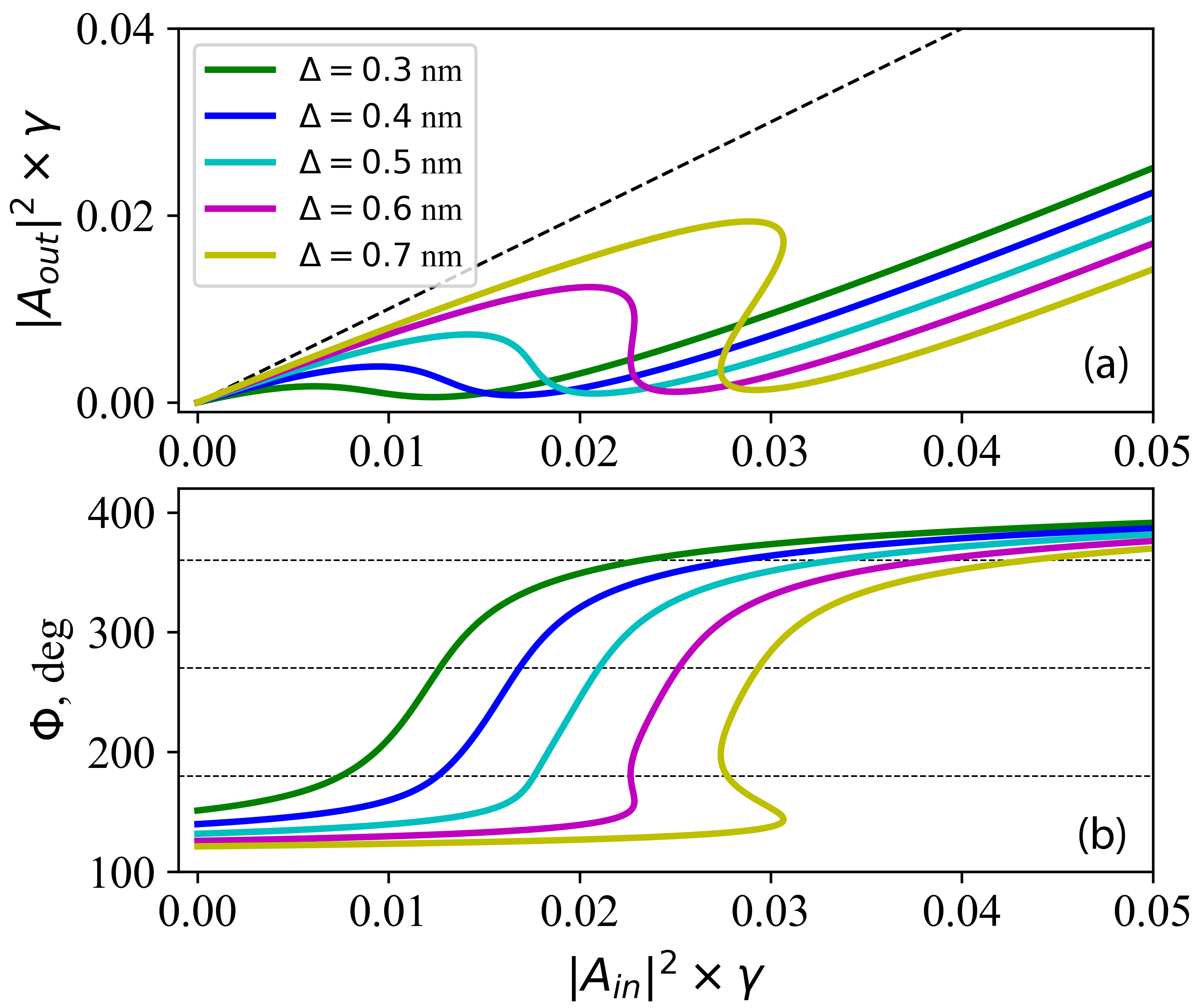}}
\caption {The dependence of output intensity (a) and phase (b) on the input intensity in nonlinear ring resonators. Different values of detuning are considered.
\label{fig2_rrnonlin_charact}}
\end{figure}

In Fig.~\ref{fig2_rrnonlin_charact} we show the dependence of phase and output intensity on the input intensity for a nonlinear ring based filter.
To keep the analysis general, we express the intensities in relative units $\gamma|A|^2$. It allows to compare different phase switching mechanisms without going much into the details of a specific nonlinear mechanism involved. At the end of the manuscript we provide the numerical simulation results with absolute intensity scale for Kerr nonlinearity in silicon.
The wavelength In Fig.~\ref{fig2_rrnonlin_charact} is slightly shifted with respect to the resonance. Different values of detuning $\Delta$ are considered (see the legend). The back dashed line denotes the  $|E_{in}|^2=|E_{out}|^2$ condition, that one would expect to observe in the linear out-of-resonance regime. Thus, it defines the asymptotics at small intensities. Increasing the input intensity we force the system to pass through the resonance. Intensity goes down and then goes up, at the same time the phase switching takes place. For larger detunings the threshold intensity grows as well. Large intensity, in turn, initiates bistable regimes (magenta and yellow curves).
The phase profiles shown in Fig~\ref{fig2_rrnonlin_charact}b are, in principle, suitable to represent the activation function. But, to obtain very sharp steps extremely high quality resonators are required. Moreover, dynamical control of sharpness can be also problematic in this case.
It is also necessary to stress that for applications in classical optics based Ising machines, that we have in mind, bistability is rather undesired effect, which we would try to avoid.  At the same time, other approaches exist, where bistability is required \cite{tezak-2020}.

\begin{figure}
\centerline{\includegraphics[width=0.5\textwidth]{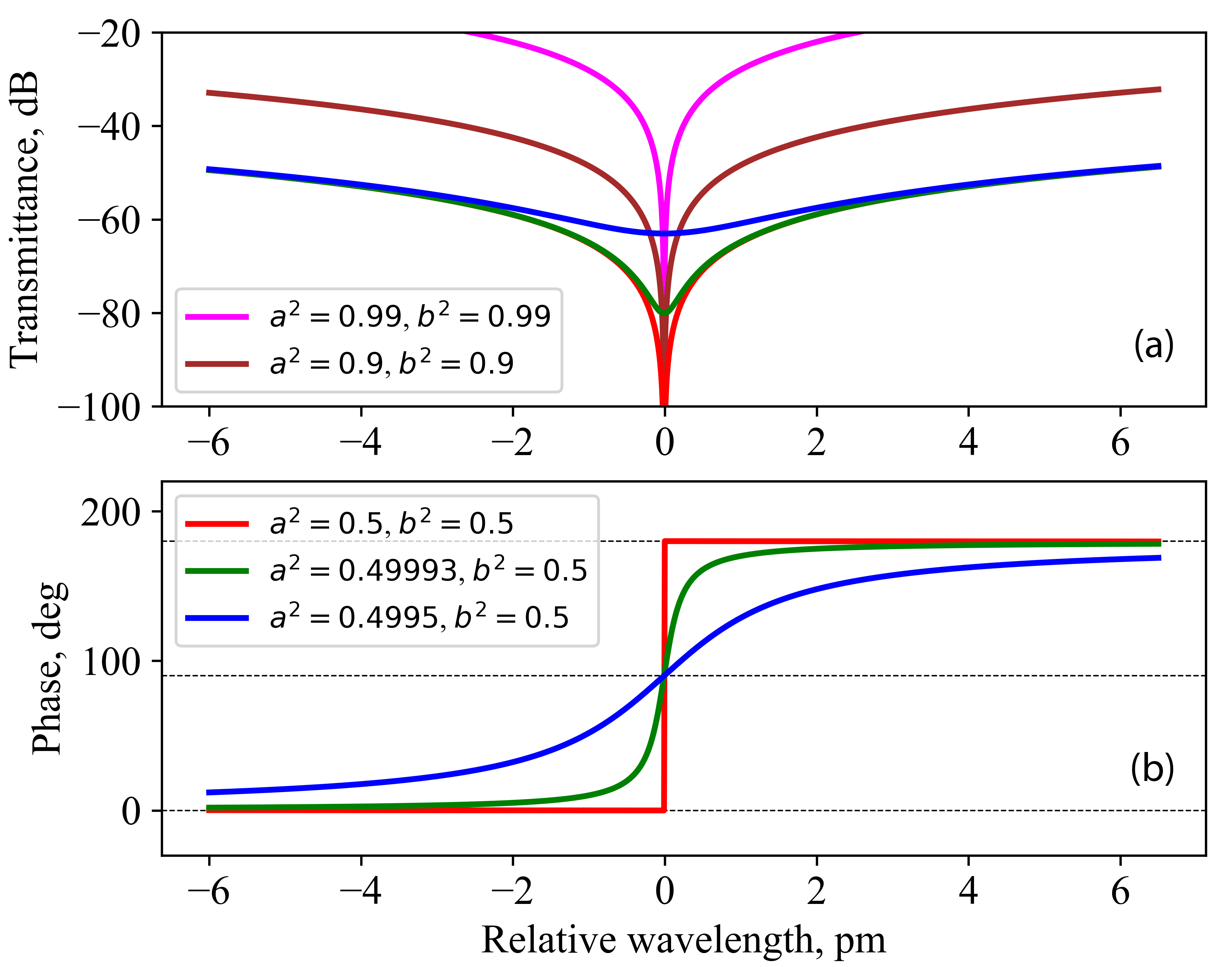}}
\caption {Transmittance and phase at the output of the resonator plotted as a function of relative wavelength close to the critical coupling regime.
\label{fig3_rrlin_cc}}
\end{figure}

Now we discuss an alternative switching scheme in the vicinity of the critical coupling regime. Basically, we are going to use the red and blue curves instead of the green curve in Fig.~\ref{fig1_rrlinphases}. The dependence of transmittance and phase in this regimes on the wavelength is shown in Fig.~\ref{fig3_rrlin_cc}. Since the plotted feature is rather thin, a relative with respect to the position of the resonance wavelength scale is used. The width, that is just few picometers in size, allows to use small values of detuning and require much smaller intensities for switching (discussed below).

Red, green and blue curves in Fig.~\ref{fig3_rrlin_cc} correspond to the lowest quality resonator with slightly varied losses. When one increases losses (by decreasing $a$) a perfectly sharp step function turns into a smoothed Fermi-Dirac-like shape, providing exactly the type of behavior we are looking for. Such a modification of losses can be achieved, for example, by integrating an absorption modulator into the ring \cite{pshenichnyuk-2019, pshenichnyuk-2019b, pshenichnyuk-2021,zemtsov-2023}. 
For the comparison two more critically coupled resonators with higher quality factors (brown and magenta curves) are considered. They both provide the perfect step functions with respect to a phase, that are indistinguishable from the red curve in Fig.~\ref{fig3_rrlin_cc}b. The difference takes place in the transmittance (Fig.~\ref{fig3_rrlin_cc}a). Since the effect is observed close to the resonance, the transmittance is in general low and we use the logarithmic scale. In the center of the resonance, where $T=0$ the curve goes to $-\infty$ in the critical coupling regime, but grows when the losses are increased (green and blue lines). In general, resonances with a higher quality provide a larger transmittance and the usage of high quality resonators is still desirable. But the quality has no influence on the curvature of the phase step.
Low transmittance also takes place for thresholders depicted in Fig.~\ref{fig2_rrnonlin_charact}. For the switching mechanism based on the critical coupling it is obviously lower, but still quite measurable. For practical applications, probably, amplifiers have to be considered. At the same time, it is important to note, that, since the minima in transmittance is an interference effect, the energy is not lost and can be collected, for example, if we add the second waveguide to the ring (drop port). This signal can be used elsewhere in the circuit, probably, even for the amplification of the main signal. Thus, despite the fact that the output transmittance is low, the overall losses in such a device can be made minimal.

\begin{figure}
\centerline{\includegraphics[width=0.5\textwidth]{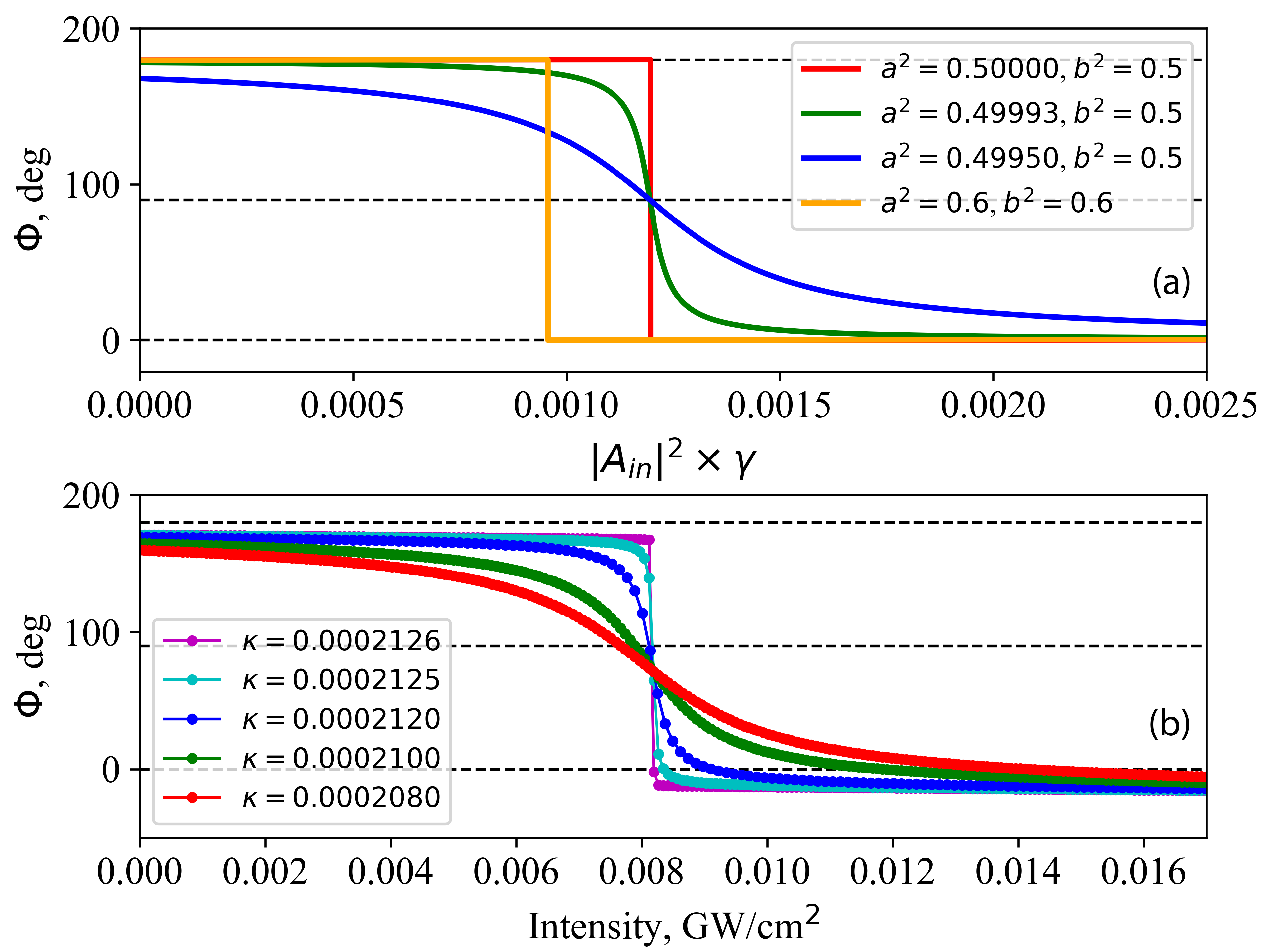}}
\caption {(a) An output phase of a nonlinear resonator as a function of input intensity obtained from the semi-analytical model. (b) Numerical results obtained from Maxwell simulations for a resonator with a variable loss coefficient $\kappa$. 
\label{fig4_fig2_rrnonlin_phase}}
\end{figure}

The phase steps produced by nonlinear resonators with a variable input intensity in a vicinity of the critical coupling regime are shown in Fig.~\ref{fig4_fig2_rrnonlin_phase}. The results extracted from the analytical computations (Eqs.~\ref{rr_nonlin_smatrix1}-\ref{rr_nonlin_smatrix2}) are shown in Fig.~\ref{fig4_fig2_rrnonlin_phase}a. 
As expected, the sharp step is observed at the critical coupling regime and it can be smeared by increasing losses in the ring. At a small value of detuning $\Delta = 6$ pm the step appears at relative intensity close to $0.0012$ which is almost $20$ times smaller to compare with the result depicted in Fig.~\ref{fig2_rrnonlin_charact}b. It allows to realize the switching at small intensities and avoid bistable regimes. 
For the comparison, one additional curve is added with a slightly increased quality factor (yellow curve in Fig.~\ref{fig4_fig2_rrnonlin_phase}a). The sharpness of the step is exactly the same, but it appears at even smaller intensity. The explanation is related to the fact that resonators with larger quality factors allow to accumulate larger intensity inside at fixed input intensity. It allows to decrease the switching threshold even further. This is another example of an advantage provided by high quality resonators, but, again, it has no influence on the quality of the thresholder.
Horizontal dashed lines in Fig.~\ref{fig4_fig2_rrnonlin_phase} are placed equidistantly each $90$ degrees. It is evident that the desired $\pi$ phase shift is obtained for the proposed device. In contrast with Fig.~\ref{fig2_rrnonlin_charact} where the shift is larger than $\pi$. It is the consequence of the fact that the system only partially passes through the broad resonance in a given intensity range, otherwise $2\pi$ shift would be obtained (see Fig.~\ref{fig1_rrlinphases} green curve). To obtain exactly $\pi$ shift (useful for applications) for the devices depicted in Fig.~\ref{fig2_rrnonlin_charact}, the detuning and intensity have to be additionally calibrated.

In Fig.~\ref{fig4_fig2_rrnonlin_phase}b we present the results of numerical simulations where nonlinear Maxwell equations were solved for a ring resonator with a radius $5$ $\mu$m and a waveguide width $220$ nm. Its actual quality factor is close to $4118$. The value of detuning $\Delta = 10$ pm is used in simulations.
Parameters $a$ and $b$ can not be directly incorporated into the numerical model. The losses are controlled through the imaginary part $\kappa$ of the refractive index (shown by different colors in Fig.~\ref{fig4_fig2_rrnonlin_phase}b). While the coupling depends on the distance between the waveguide and the ring, which is equal to $239$ nm in the simulation. We also incorporate explicitly the Kerr coefficient for silicon $n_2 = 6\times 10^{-5}$ cm$^2$/GW.
The model is quite different from the analytical one for many reasons. For example, the correct dependence of the effective index on the wavelength is taken into account. The coupler is an extensive structure instead of 'point coupler' considered in analytical computations. Nonlinearity takes place not only in the ring, but also in the straight part of the waveguide.
Nevertheless, the results of both approaches are quite similar. In the numerical model an absolute intensity scale linked directly with the Kerr nonlinearity of silicon is obtained. Usually, to observe this type of nonlinearity in silicon, much larger intensities should be applied, that are achievable only for pulsed lasers. In our example the threshold intensity is rather small ($0.008$ GW/cm$^2$, instead of usual in this context few GW/cm$^2$ \cite{rukhlenko-2010}). Thus, the suggested mechanism may be practically applied even for continuous lasers. Kerr nonlinearity in silicon is also quite fast and its usage has many advantages for applications.


Summarizing, we suggest to utilize integrated optical resonators to obtain sharp intensity dependent $\pi$ phase shifts. The mechanism is based on the nonlinear switching in a vicinity of the critical coupling regime. The sharpness is controllable and it is not confined by the quality factors of resonators. The described behavior is perfect for the realization of activation functions in neuromorphic photonic chips. High quality factors are still advantageous for two reasons: larger transmittance and smaller activation thresholds in the nonlinear regime, but they do not influence the quality of the function itself.
Having in hand such a resonator one can build it in into an arm of MZI, like, for example, described in literature \cite{huang-2021, jha-2020, huang-2019}. With a major difference, that along with a heater, that can be used to shift resonances and modify the positions of steps, the ring should contain a controllable source of losses, like, for example, an amplitude modulator. It should be calibrated to work near the critical coupling regime. Extremely sharp steps allow to use small switching intensities. This, in turn, allows to use fast Kerr nonlinearity in silicon, instead of free carrier effects \cite{borghi-2021}. It also allows to use a continuous laser regime, instead of pulses.
Taking into account the robustness of the considered mechanism with respect to quality factors, probably, other types of resonators also deserve attention for future research in this direction, like, the samples based on Bragg mirrors, photonic crystals and inverse design structures \cite{kim-2024,pruessner-2007}. Such studies would potentially allow to make thresholders much more compact compared to ring resonators.

\bibliography{paper_afunction}

\begin{thebibliography}{32}%
\makeatletter
\providecommand \@ifxundefined [1]{%
 \@ifx{#1\undefined}
}%
\providecommand \@ifnum [1]{%
 \ifnum #1\expandafter \@firstoftwo
 \else \expandafter \@secondoftwo
 \fi
}%
\providecommand \@ifx [1]{%
 \ifx #1\expandafter \@firstoftwo
 \else \expandafter \@secondoftwo
 \fi
}%
\providecommand \natexlab [1]{#1}%
\providecommand \enquote  [1]{``#1''}%
\providecommand \bibnamefont  [1]{#1}%
\providecommand \bibfnamefont [1]{#1}%
\providecommand \citenamefont [1]{#1}%
\providecommand \href@noop [0]{\@secondoftwo}%
\providecommand \href [0]{\begingroup \@sanitize@url \@href}%
\providecommand \@href[1]{\@@startlink{#1}\@@href}%
\providecommand \@@href[1]{\endgroup#1\@@endlink}%
\providecommand \@sanitize@url [0]{\catcode `\\12\catcode `\$12\catcode
  `\&12\catcode `\#12\catcode `\^12\catcode `\_12\catcode `\%12\relax}%
\providecommand \@@startlink[1]{}%
\providecommand \@@endlink[0]{}%
\providecommand \url  [0]{\begingroup\@sanitize@url \@url }%
\providecommand \@url [1]{\endgroup\@href {#1}{\urlprefix }}%
\providecommand \urlprefix  [0]{URL }%
\providecommand \Eprint [0]{\href }%
\providecommand \doibase [0]{http://dx.doi.org/}%
\providecommand \selectlanguage [0]{\@gobble}%
\providecommand \bibinfo  [0]{\@secondoftwo}%
\providecommand \bibfield  [0]{\@secondoftwo}%
\providecommand \translation [1]{[#1]}%
\providecommand \BibitemOpen [0]{}%
\providecommand \bibitemStop [0]{}%
\providecommand \bibitemNoStop [0]{.\EOS\space}%
\providecommand \EOS [0]{\spacefactor3000\relax}%
\providecommand \BibitemShut  [1]{\csname bibitem#1\endcsname}%
\let\auto@bib@innerbib\@empty
\bibitem [{\citenamefont {Bogaerts}\ \emph {et~al.}(2012)\citenamefont
  {Bogaerts}, \citenamefont {De~Heyn}, \citenamefont {Van~Vaerenbergh},
  \citenamefont {De~Vos}, \citenamefont {Kumar~Selvaraja}, \citenamefont
  {Claes}, \citenamefont {Dumon}, \citenamefont {Bienstman}, \citenamefont
  {Van~Thourhout},\ and\ \citenamefont {Baets}}]{bogaerts-2012}%
  \BibitemOpen
  \bibfield  {author} {\bibinfo {author} {\bibfnamefont {W.}~\bibnamefont
  {Bogaerts}}, \bibinfo {author} {\bibfnamefont {P.}~\bibnamefont {De~Heyn}},
  \bibinfo {author} {\bibfnamefont {T.}~\bibnamefont {Van~Vaerenbergh}},
  \bibinfo {author} {\bibfnamefont {K.}~\bibnamefont {De~Vos}}, \bibinfo
  {author} {\bibfnamefont {S.}~\bibnamefont {Kumar~Selvaraja}}, \bibinfo
  {author} {\bibfnamefont {T.}~\bibnamefont {Claes}}, \bibinfo {author}
  {\bibfnamefont {P.}~\bibnamefont {Dumon}}, \bibinfo {author} {\bibfnamefont
  {P.}~\bibnamefont {Bienstman}}, \bibinfo {author} {\bibfnamefont
  {D.}~\bibnamefont {Van~Thourhout}}, \ and\ \bibinfo {author} {\bibfnamefont
  {R.}~\bibnamefont {Baets}},\ }\href@noop {} {\bibfield  {journal} {\bibinfo
  {journal} {Laser \& Photonics Reviews}\ }\textbf {\bibinfo {volume} {6}},\
  \bibinfo {pages} {47} (\bibinfo {year} {2012})}\BibitemShut {NoStop}%
\bibitem [{\citenamefont {Yariv}\ \emph {et~al.}(2007)\citenamefont {Yariv},
  \citenamefont {Yeh},\ and\ \citenamefont {Yariv}}]{yariv}%
  \BibitemOpen
  \bibfield  {author} {\bibinfo {author} {\bibfnamefont {A.}~\bibnamefont
  {Yariv}}, \bibinfo {author} {\bibfnamefont {P.}~\bibnamefont {Yeh}}, \ and\
  \bibinfo {author} {\bibfnamefont {A.}~\bibnamefont {Yariv}},\ }\href@noop {}
  {\emph {\bibinfo {title} {Photonics: optical electronics in modern
  communications}}},\ Vol.~\bibinfo {volume} {6}\ (\bibinfo  {publisher}
  {Oxford university press New York},\ \bibinfo {year} {2007})\BibitemShut
  {NoStop}%
\bibitem [{\citenamefont {Rukhlenko}\ \emph {et~al.}(2010)\citenamefont
  {Rukhlenko}, \citenamefont {Premaratne},\ and\ \citenamefont
  {Agrawal}}]{rukhlenko-2010}%
  \BibitemOpen
  \bibfield  {author} {\bibinfo {author} {\bibfnamefont {I.~D.}\ \bibnamefont
  {Rukhlenko}}, \bibinfo {author} {\bibfnamefont {M.}~\bibnamefont
  {Premaratne}}, \ and\ \bibinfo {author} {\bibfnamefont {G.~P.}\ \bibnamefont
  {Agrawal}},\ }\href@noop {} {\bibfield  {journal} {\bibinfo  {journal}
  {Optics letters}\ }\textbf {\bibinfo {volume} {35}},\ \bibinfo {pages} {55}
  (\bibinfo {year} {2010})}\BibitemShut {NoStop}%
\bibitem [{\citenamefont {Chen}\ \emph {et~al.}(2012)\citenamefont {Chen},
  \citenamefont {Badioli}, \citenamefont {Alonso-Gonz{\'a}lez}, \citenamefont
  {Thongrattanasiri}, \citenamefont {Huth}, \citenamefont {Osmond},
  \citenamefont {Spasenovi{\'c}}, \citenamefont {Centeno}, \citenamefont
  {Pesquera}, \citenamefont {Godignon} \emph {et~al.}}]{chen-2012}%
  \BibitemOpen
  \bibfield  {author} {\bibinfo {author} {\bibfnamefont {J.}~\bibnamefont
  {Chen}}, \bibinfo {author} {\bibfnamefont {M.}~\bibnamefont {Badioli}},
  \bibinfo {author} {\bibfnamefont {P.}~\bibnamefont {Alonso-Gonz{\'a}lez}},
  \bibinfo {author} {\bibfnamefont {S.}~\bibnamefont {Thongrattanasiri}},
  \bibinfo {author} {\bibfnamefont {F.}~\bibnamefont {Huth}}, \bibinfo {author}
  {\bibfnamefont {J.}~\bibnamefont {Osmond}}, \bibinfo {author} {\bibfnamefont
  {M.}~\bibnamefont {Spasenovi{\'c}}}, \bibinfo {author} {\bibfnamefont
  {A.}~\bibnamefont {Centeno}}, \bibinfo {author} {\bibfnamefont
  {A.}~\bibnamefont {Pesquera}}, \bibinfo {author} {\bibfnamefont
  {P.}~\bibnamefont {Godignon}},  \emph {et~al.},\ }\href@noop {} {\bibfield
  {journal} {\bibinfo  {journal} {Nature}\ }\textbf {\bibinfo {volume} {487}},\
  \bibinfo {pages} {77} (\bibinfo {year} {2012})}\BibitemShut {NoStop}%
\bibitem [{\citenamefont {Huang}\ \emph {et~al.}(2019)\citenamefont {Huang},
  \citenamefont {De~Lima}, \citenamefont {Jha}, \citenamefont {Abbaslou},
  \citenamefont {Tait}, \citenamefont {Shastri},\ and\ \citenamefont
  {Prucnal}}]{huang-2019}%
  \BibitemOpen
  \bibfield  {author} {\bibinfo {author} {\bibfnamefont {C.}~\bibnamefont
  {Huang}}, \bibinfo {author} {\bibfnamefont {T.~F.}\ \bibnamefont {De~Lima}},
  \bibinfo {author} {\bibfnamefont {A.}~\bibnamefont {Jha}}, \bibinfo {author}
  {\bibfnamefont {S.}~\bibnamefont {Abbaslou}}, \bibinfo {author}
  {\bibfnamefont {A.~N.}\ \bibnamefont {Tait}}, \bibinfo {author}
  {\bibfnamefont {B.~J.}\ \bibnamefont {Shastri}}, \ and\ \bibinfo {author}
  {\bibfnamefont {P.~R.}\ \bibnamefont {Prucnal}},\ }\href@noop {} {\bibfield
  {journal} {\bibinfo  {journal} {IEEE Photonics Technology Letters}\ }\textbf
  {\bibinfo {volume} {31}},\ \bibinfo {pages} {1834} (\bibinfo {year}
  {2019})}\BibitemShut {NoStop}%
\bibitem [{\citenamefont {Xu}\ and\ \citenamefont {Lipson}(2007)}]{xu-2007}%
  \BibitemOpen
  \bibfield  {author} {\bibinfo {author} {\bibfnamefont {Q.}~\bibnamefont
  {Xu}}\ and\ \bibinfo {author} {\bibfnamefont {M.}~\bibnamefont {Lipson}},\
  }\href@noop {} {\bibfield  {journal} {\bibinfo  {journal} {Optics express}\
  }\textbf {\bibinfo {volume} {15}},\ \bibinfo {pages} {924} (\bibinfo {year}
  {2007})}\BibitemShut {NoStop}%
\bibitem [{\citenamefont {Tezak}\ \emph {et~al.}(2020)\citenamefont {Tezak},
  \citenamefont {Van~Vaerenbergh}, \citenamefont {Pelc}, \citenamefont
  {Mendoza}, \citenamefont {Kielpinski}, \citenamefont {Mabuchi},\ and\
  \citenamefont {Beausoleil}}]{tezak-2020}%
  \BibitemOpen
  \bibfield  {author} {\bibinfo {author} {\bibfnamefont {N.}~\bibnamefont
  {Tezak}}, \bibinfo {author} {\bibfnamefont {T.}~\bibnamefont
  {Van~Vaerenbergh}}, \bibinfo {author} {\bibfnamefont {J.~S.}\ \bibnamefont
  {Pelc}}, \bibinfo {author} {\bibfnamefont {G.~J.}\ \bibnamefont {Mendoza}},
  \bibinfo {author} {\bibfnamefont {D.}~\bibnamefont {Kielpinski}}, \bibinfo
  {author} {\bibfnamefont {H.}~\bibnamefont {Mabuchi}}, \ and\ \bibinfo
  {author} {\bibfnamefont {R.~G.}\ \bibnamefont {Beausoleil}},\ }\href@noop {}
  {\bibfield  {journal} {\bibinfo  {journal} {IEEE Journal of Selected Topics
  in Quantum Electronics}\ }\textbf {\bibinfo {volume} {26}},\ \bibinfo {pages}
  {1} (\bibinfo {year} {2020})}\BibitemShut {NoStop}%
\bibitem [{\citenamefont {Tezak}\ and\ \citenamefont
  {Mabuchi}(2015)}]{tezak-2015}%
  \BibitemOpen
  \bibfield  {author} {\bibinfo {author} {\bibfnamefont {N.}~\bibnamefont
  {Tezak}}\ and\ \bibinfo {author} {\bibfnamefont {H.}~\bibnamefont
  {Mabuchi}},\ }\href@noop {} {\bibfield  {journal} {\bibinfo  {journal} {EPJ
  Quantum Technology}\ }\textbf {\bibinfo {volume} {2}},\ \bibinfo {pages} {1}
  (\bibinfo {year} {2015})}\BibitemShut {NoStop}%
\bibitem [{\citenamefont {Huang}\ \emph {et~al.}(2021)\citenamefont {Huang},
  \citenamefont {Jha}, \citenamefont {De~Lima}, \citenamefont {Tait},
  \citenamefont {Shastri},\ and\ \citenamefont {Prucnal}}]{huang-2021}%
  \BibitemOpen
  \bibfield  {author} {\bibinfo {author} {\bibfnamefont {C.}~\bibnamefont
  {Huang}}, \bibinfo {author} {\bibfnamefont {A.}~\bibnamefont {Jha}}, \bibinfo
  {author} {\bibfnamefont {T.~F.}\ \bibnamefont {De~Lima}}, \bibinfo {author}
  {\bibfnamefont {A.~N.}\ \bibnamefont {Tait}}, \bibinfo {author}
  {\bibfnamefont {B.~J.}\ \bibnamefont {Shastri}}, \ and\ \bibinfo {author}
  {\bibfnamefont {P.~R.}\ \bibnamefont {Prucnal}},\ }\href@noop {} {\bibfield
  {journal} {\bibinfo  {journal} {IEEE Journal of selected topics in quantum
  electronics}\ }\textbf {\bibinfo {volume} {27}},\ \bibinfo {pages} {1}
  (\bibinfo {year} {2021})}\BibitemShut {NoStop}%
\bibitem [{\citenamefont {Hertz}(2018)}]{hertz}%
  \BibitemOpen
  \bibfield  {author} {\bibinfo {author} {\bibfnamefont {J.~A.}\ \bibnamefont
  {Hertz}},\ }\href@noop {} {\emph {\bibinfo {title} {Introduction to the
  theory of neural computation}}}\ (\bibinfo  {publisher} {Crc Press},\
  \bibinfo {year} {2018})\BibitemShut {NoStop}%
\bibitem [{\citenamefont {Prucnal}\ and\ \citenamefont
  {Shastri}(2017)}]{prucnal}%
  \BibitemOpen
  \bibfield  {author} {\bibinfo {author} {\bibfnamefont {P.~R.}\ \bibnamefont
  {Prucnal}}\ and\ \bibinfo {author} {\bibfnamefont {B.~J.}\ \bibnamefont
  {Shastri}},\ }\href@noop {} {\emph {\bibinfo {title} {Neuromorphic
  photonics}}}\ (\bibinfo  {publisher} {CRC press},\ \bibinfo {year}
  {2017})\BibitemShut {NoStop}%
\bibitem [{\citenamefont {Ertel}(2017)}]{ertel}%
  \BibitemOpen
  \bibfield  {author} {\bibinfo {author} {\bibfnamefont {W.}~\bibnamefont
  {Ertel}},\ }\href@noop {} {\emph {\bibinfo {title} {Introduction to
  Artificial Intelligence}}}\ (\bibinfo  {publisher} {Springer International
  Publishing AG},\ \bibinfo {year} {2017})\BibitemShut {NoStop}%
\bibitem [{\citenamefont {Kielpinski}\ \emph {et~al.}(2016)\citenamefont
  {Kielpinski}, \citenamefont {Bose}, \citenamefont {Pelc}, \citenamefont
  {Van~Vaerenbergh}, \citenamefont {Mendoza}, \citenamefont {Tezak},\ and\
  \citenamefont {Beausoleil}}]{kielpinski-2016}%
  \BibitemOpen
  \bibfield  {author} {\bibinfo {author} {\bibfnamefont {D.}~\bibnamefont
  {Kielpinski}}, \bibinfo {author} {\bibfnamefont {R.}~\bibnamefont {Bose}},
  \bibinfo {author} {\bibfnamefont {J.}~\bibnamefont {Pelc}}, \bibinfo {author}
  {\bibfnamefont {T.}~\bibnamefont {Van~Vaerenbergh}}, \bibinfo {author}
  {\bibfnamefont {G.}~\bibnamefont {Mendoza}}, \bibinfo {author} {\bibfnamefont
  {N.}~\bibnamefont {Tezak}}, \ and\ \bibinfo {author} {\bibfnamefont {R.~G.}\
  \bibnamefont {Beausoleil}},\ }in\ \href@noop {} {\emph {\bibinfo {booktitle}
  {2016 IEEE International conference on rebooting computing (ICRC)}}}\
  (\bibinfo {organization} {IEEE},\ \bibinfo {year} {2016})\ pp.\ \bibinfo
  {pages} {1--4}\BibitemShut {NoStop}%
\bibitem [{\citenamefont {Newman}\ and\ \citenamefont
  {Barkema}(1999)}]{newman}%
  \BibitemOpen
  \bibfield  {author} {\bibinfo {author} {\bibfnamefont {M.~E.}\ \bibnamefont
  {Newman}}\ and\ \bibinfo {author} {\bibfnamefont {G.~T.}\ \bibnamefont
  {Barkema}},\ }\href@noop {} {\emph {\bibinfo {title} {Monte Carlo methods in
  statistical physics}}}\ (\bibinfo  {publisher} {Clarendon Press},\ \bibinfo
  {year} {1999})\BibitemShut {NoStop}%
\bibitem [{\citenamefont {Campo}\ and\ \citenamefont
  {P{\'e}rez-L{\'o}pez}(2022)}]{campo-2022}%
  \BibitemOpen
  \bibfield  {author} {\bibinfo {author} {\bibfnamefont {J.~R.~R.}\
  \bibnamefont {Campo}}\ and\ \bibinfo {author} {\bibfnamefont
  {D.}~\bibnamefont {P{\'e}rez-L{\'o}pez}},\ }\href@noop {} {\bibfield
  {journal} {\bibinfo  {journal} {IEEE Journal of Selected Topics in Quantum
  Electronics}\ }\textbf {\bibinfo {volume} {28}},\ \bibinfo {pages} {1}
  (\bibinfo {year} {2022})}\BibitemShut {NoStop}%
\bibitem [{\citenamefont {Pshenichnyuk}\ \emph {et~al.}(2024)\citenamefont
  {Pshenichnyuk}, \citenamefont {Yousry}, \citenamefont {Zemtsov},
  \citenamefont {Kosolobov},\ and\ \citenamefont
  {Drachev}}]{pshenichnyuk-2024}%
  \BibitemOpen
  \bibfield  {author} {\bibinfo {author} {\bibfnamefont {I.~A.}\ \bibnamefont
  {Pshenichnyuk}}, \bibinfo {author} {\bibfnamefont {F.}~\bibnamefont
  {Yousry}}, \bibinfo {author} {\bibfnamefont {D.~S.}\ \bibnamefont {Zemtsov}},
  \bibinfo {author} {\bibfnamefont {S.~S.}\ \bibnamefont {Kosolobov}}, \ and\
  \bibinfo {author} {\bibfnamefont {V.~P.}\ \bibnamefont {Drachev}},\
  }\href@noop {} {\bibfield  {journal} {\bibinfo  {journal} {Physical Review
  B}\ }\textbf {\bibinfo {volume} {109}},\ \bibinfo {pages} {035401} (\bibinfo
  {year} {2024})}\BibitemShut {NoStop}%
\bibitem [{\citenamefont {Heebner}\ \emph {et~al.}(2004)\citenamefont
  {Heebner}, \citenamefont {Wong}, \citenamefont {Schweinsberg}, \citenamefont
  {Boyd},\ and\ \citenamefont {Jackson}}]{heebner-2004}%
  \BibitemOpen
  \bibfield  {author} {\bibinfo {author} {\bibfnamefont {J.~E.}\ \bibnamefont
  {Heebner}}, \bibinfo {author} {\bibfnamefont {V.}~\bibnamefont {Wong}},
  \bibinfo {author} {\bibfnamefont {A.}~\bibnamefont {Schweinsberg}}, \bibinfo
  {author} {\bibfnamefont {R.~W.}\ \bibnamefont {Boyd}}, \ and\ \bibinfo
  {author} {\bibfnamefont {D.~J.}\ \bibnamefont {Jackson}},\ }\href@noop {}
  {\bibfield  {journal} {\bibinfo  {journal} {IEEE journal of quantum
  electronics}\ }\textbf {\bibinfo {volume} {40}},\ \bibinfo {pages} {726}
  (\bibinfo {year} {2004})}\BibitemShut {NoStop}%
\bibitem [{\citenamefont {Liu}\ \emph {et~al.}(2008)\citenamefont {Liu},
  \citenamefont {Wang}, \citenamefont {Qiang}, \citenamefont {Ye},
  \citenamefont {Zhang}, \citenamefont {Qiu},\ and\ \citenamefont
  {Su}}]{liu-2008}%
  \BibitemOpen
  \bibfield  {author} {\bibinfo {author} {\bibfnamefont {F.}~\bibnamefont
  {Liu}}, \bibinfo {author} {\bibfnamefont {T.}~\bibnamefont {Wang}}, \bibinfo
  {author} {\bibfnamefont {L.}~\bibnamefont {Qiang}}, \bibinfo {author}
  {\bibfnamefont {T.}~\bibnamefont {Ye}}, \bibinfo {author} {\bibfnamefont
  {Z.}~\bibnamefont {Zhang}}, \bibinfo {author} {\bibfnamefont
  {M.}~\bibnamefont {Qiu}}, \ and\ \bibinfo {author} {\bibfnamefont
  {Y.}~\bibnamefont {Su}},\ }\href@noop {} {\bibfield  {journal} {\bibinfo
  {journal} {Optics Express}\ }\textbf {\bibinfo {volume} {16}},\ \bibinfo
  {pages} {15880} (\bibinfo {year} {2008})}\BibitemShut {NoStop}%
\bibitem [{\citenamefont {Chalupnik}\ \emph {et~al.}(2023)\citenamefont
  {Chalupnik}, \citenamefont {Singh}, \citenamefont {Leatham}, \citenamefont
  {Lon{\v{c}}ar},\ and\ \citenamefont {Soltani}}]{chalupnik-2023}%
  \BibitemOpen
  \bibfield  {author} {\bibinfo {author} {\bibfnamefont {M.}~\bibnamefont
  {Chalupnik}}, \bibinfo {author} {\bibfnamefont {A.}~\bibnamefont {Singh}},
  \bibinfo {author} {\bibfnamefont {J.}~\bibnamefont {Leatham}}, \bibinfo
  {author} {\bibfnamefont {M.}~\bibnamefont {Lon{\v{c}}ar}}, \ and\ \bibinfo
  {author} {\bibfnamefont {M.}~\bibnamefont {Soltani}},\ }\href@noop {}
  {\bibfield  {journal} {\bibinfo  {journal} {APL Photonics}\ }\textbf
  {\bibinfo {volume} {8}} (\bibinfo {year} {2023})}\BibitemShut {NoStop}%
\bibitem [{\citenamefont {Shen}\ \emph {et~al.}(2017)\citenamefont {Shen},
  \citenamefont {Harris}, \citenamefont {Skirlo}, \citenamefont {Prabhu},
  \citenamefont {Baehr-Jones}, \citenamefont {Hochberg}, \citenamefont {Sun},
  \citenamefont {Zhao}, \citenamefont {Larochelle}, \citenamefont {Englund}
  \emph {et~al.}}]{shen-2017}%
  \BibitemOpen
  \bibfield  {author} {\bibinfo {author} {\bibfnamefont {Y.}~\bibnamefont
  {Shen}}, \bibinfo {author} {\bibfnamefont {N.~C.}\ \bibnamefont {Harris}},
  \bibinfo {author} {\bibfnamefont {S.}~\bibnamefont {Skirlo}}, \bibinfo
  {author} {\bibfnamefont {M.}~\bibnamefont {Prabhu}}, \bibinfo {author}
  {\bibfnamefont {T.}~\bibnamefont {Baehr-Jones}}, \bibinfo {author}
  {\bibfnamefont {M.}~\bibnamefont {Hochberg}}, \bibinfo {author}
  {\bibfnamefont {X.}~\bibnamefont {Sun}}, \bibinfo {author} {\bibfnamefont
  {S.}~\bibnamefont {Zhao}}, \bibinfo {author} {\bibfnamefont {H.}~\bibnamefont
  {Larochelle}}, \bibinfo {author} {\bibfnamefont {D.}~\bibnamefont {Englund}},
   \emph {et~al.},\ }\href@noop {} {\bibfield  {journal} {\bibinfo  {journal}
  {Nature photonics}\ }\textbf {\bibinfo {volume} {11}},\ \bibinfo {pages}
  {441} (\bibinfo {year} {2017})}\BibitemShut {NoStop}%
\bibitem [{\citenamefont {Harris}\ \emph {et~al.}(2018)\citenamefont {Harris},
  \citenamefont {Carolan}, \citenamefont {Bunandar}, \citenamefont {Prabhu},
  \citenamefont {Hochberg}, \citenamefont {Baehr-Jones}, \citenamefont {Fanto},
  \citenamefont {Smith}, \citenamefont {Tison}, \citenamefont {Alsing} \emph
  {et~al.}}]{harris-2018}%
  \BibitemOpen
  \bibfield  {author} {\bibinfo {author} {\bibfnamefont {N.~C.}\ \bibnamefont
  {Harris}}, \bibinfo {author} {\bibfnamefont {J.}~\bibnamefont {Carolan}},
  \bibinfo {author} {\bibfnamefont {D.}~\bibnamefont {Bunandar}}, \bibinfo
  {author} {\bibfnamefont {M.}~\bibnamefont {Prabhu}}, \bibinfo {author}
  {\bibfnamefont {M.}~\bibnamefont {Hochberg}}, \bibinfo {author}
  {\bibfnamefont {T.}~\bibnamefont {Baehr-Jones}}, \bibinfo {author}
  {\bibfnamefont {M.~L.}\ \bibnamefont {Fanto}}, \bibinfo {author}
  {\bibfnamefont {A.~M.}\ \bibnamefont {Smith}}, \bibinfo {author}
  {\bibfnamefont {C.~C.}\ \bibnamefont {Tison}}, \bibinfo {author}
  {\bibfnamefont {P.~M.}\ \bibnamefont {Alsing}},  \emph {et~al.},\ }\href@noop
  {} {\bibfield  {journal} {\bibinfo  {journal} {Optica}\ }\textbf {\bibinfo
  {volume} {5}},\ \bibinfo {pages} {1623} (\bibinfo {year} {2018})}\BibitemShut
  {NoStop}%
\bibitem [{\citenamefont {Burla}\ \emph {et~al.}(2015)\citenamefont {Burla},
  \citenamefont {Crockett}, \citenamefont {Chrostowski},\ and\ \citenamefont
  {Aza{\~n}a}}]{burla-2015}%
  \BibitemOpen
  \bibfield  {author} {\bibinfo {author} {\bibfnamefont {M.}~\bibnamefont
  {Burla}}, \bibinfo {author} {\bibfnamefont {B.}~\bibnamefont {Crockett}},
  \bibinfo {author} {\bibfnamefont {L.}~\bibnamefont {Chrostowski}}, \ and\
  \bibinfo {author} {\bibfnamefont {J.}~\bibnamefont {Aza{\~n}a}},\ }in\
  \href@noop {} {\emph {\bibinfo {booktitle} {2015 International Topical
  Meeting on Microwave Photonics (MWP)}}}\ (\bibinfo {organization} {IEEE},\
  \bibinfo {year} {2015})\ pp.\ \bibinfo {pages} {1--4}\BibitemShut {NoStop}%
\bibitem [{\citenamefont {Haus}(1984)}]{haus}%
  \BibitemOpen
  \bibfield  {author} {\bibinfo {author} {\bibfnamefont {H.~A.}\ \bibnamefont
  {Haus}},\ }\href@noop {} {\emph {\bibinfo {title} {Waves and fields in
  optoelectronics}}}\ (\bibinfo  {publisher} {Prentice-Hall, Inc.},\ \bibinfo
  {year} {1984})\BibitemShut {NoStop}%
\bibitem [{\citenamefont {Boyd}(2020)}]{boyd}%
  \BibitemOpen
  \bibfield  {author} {\bibinfo {author} {\bibfnamefont {R.}~\bibnamefont
  {Boyd}},\ }\href@noop {} {\emph {\bibinfo {title} {Nonlinear Optics}}}\
  (\bibinfo  {publisher} {Elsevier},\ \bibinfo {year} {2020})\BibitemShut
  {NoStop}%
\bibitem [{\citenamefont {Pshenichnyuk}\ \emph
  {et~al.}(2019{\natexlab{a}})\citenamefont {Pshenichnyuk}, \citenamefont
  {Nazarikov}, \citenamefont {Kosolobov}, \citenamefont {Maimistov},\ and\
  \citenamefont {Drachev}}]{pshenichnyuk-2019}%
  \BibitemOpen
  \bibfield  {author} {\bibinfo {author} {\bibfnamefont {I.~A.}\ \bibnamefont
  {Pshenichnyuk}}, \bibinfo {author} {\bibfnamefont {G.~I.}\ \bibnamefont
  {Nazarikov}}, \bibinfo {author} {\bibfnamefont {S.~S.}\ \bibnamefont
  {Kosolobov}}, \bibinfo {author} {\bibfnamefont {A.~I.}\ \bibnamefont
  {Maimistov}}, \ and\ \bibinfo {author} {\bibfnamefont {V.~P.}\ \bibnamefont
  {Drachev}},\ }\href@noop {} {\bibfield  {journal} {\bibinfo  {journal}
  {Physical Review B}\ }\textbf {\bibinfo {volume} {100}},\ \bibinfo {pages}
  {195434} (\bibinfo {year} {2019}{\natexlab{a}})}\BibitemShut {NoStop}%
\bibitem [{\citenamefont {Pshenichnyuk}\ \emph
  {et~al.}(2019{\natexlab{b}})\citenamefont {Pshenichnyuk}, \citenamefont
  {Kosolobov},\ and\ \citenamefont {Drachev}}]{pshenichnyuk-2019b}%
  \BibitemOpen
  \bibfield  {author} {\bibinfo {author} {\bibfnamefont {I.~A.}\ \bibnamefont
  {Pshenichnyuk}}, \bibinfo {author} {\bibfnamefont {S.~S.}\ \bibnamefont
  {Kosolobov}}, \ and\ \bibinfo {author} {\bibfnamefont {V.~P.}\ \bibnamefont
  {Drachev}},\ }\href@noop {} {\bibfield  {journal} {\bibinfo  {journal}
  {Applied Sciences}\ }\textbf {\bibinfo {volume} {9}},\ \bibinfo {pages}
  {4834} (\bibinfo {year} {2019}{\natexlab{b}})}\BibitemShut {NoStop}%
\bibitem [{\citenamefont {Pshenichnyuk}\ \emph {et~al.}(2021)\citenamefont
  {Pshenichnyuk}, \citenamefont {Kosolobov},\ and\ \citenamefont
  {Drachev}}]{pshenichnyuk-2021}%
  \BibitemOpen
  \bibfield  {author} {\bibinfo {author} {\bibfnamefont {I.~A.}\ \bibnamefont
  {Pshenichnyuk}}, \bibinfo {author} {\bibfnamefont {S.~S.}\ \bibnamefont
  {Kosolobov}}, \ and\ \bibinfo {author} {\bibfnamefont {V.~P.}\ \bibnamefont
  {Drachev}},\ }\href@noop {} {\bibfield  {journal} {\bibinfo  {journal}
  {Physical Review B}\ }\textbf {\bibinfo {volume} {103}},\ \bibinfo {pages}
  {115404} (\bibinfo {year} {2021})}\BibitemShut {NoStop}%
\bibitem [{\citenamefont {Zemtsov}\ \emph {et~al.}(2023)\citenamefont
  {Zemtsov}, \citenamefont {Pshenichnyuk}, \citenamefont {Kosolobov},
  \citenamefont {Zemtsova}, \citenamefont {Zhigunov}, \citenamefont {Smirnov},
  \citenamefont {Garbuzov},\ and\ \citenamefont {Drachev}}]{zemtsov-2023}%
  \BibitemOpen
  \bibfield  {author} {\bibinfo {author} {\bibfnamefont {D.~S.}\ \bibnamefont
  {Zemtsov}}, \bibinfo {author} {\bibfnamefont {I.~A.}\ \bibnamefont
  {Pshenichnyuk}}, \bibinfo {author} {\bibfnamefont {S.~S.}\ \bibnamefont
  {Kosolobov}}, \bibinfo {author} {\bibfnamefont {A.~K.}\ \bibnamefont
  {Zemtsova}}, \bibinfo {author} {\bibfnamefont {D.~M.}\ \bibnamefont
  {Zhigunov}}, \bibinfo {author} {\bibfnamefont {A.~S.}\ \bibnamefont
  {Smirnov}}, \bibinfo {author} {\bibfnamefont {K.~N.}\ \bibnamefont
  {Garbuzov}}, \ and\ \bibinfo {author} {\bibfnamefont {V.~P.}\ \bibnamefont
  {Drachev}},\ }\href@noop {} {\bibfield  {journal} {\bibinfo  {journal}
  {Journal of Lightwave Technology}\ } (\bibinfo {year} {2023})}\BibitemShut
  {NoStop}%
\bibitem [{\citenamefont {Jha}\ \emph {et~al.}(2020)\citenamefont {Jha},
  \citenamefont {Huang},\ and\ \citenamefont {Prucnal}}]{jha-2020}%
  \BibitemOpen
  \bibfield  {author} {\bibinfo {author} {\bibfnamefont {A.}~\bibnamefont
  {Jha}}, \bibinfo {author} {\bibfnamefont {C.}~\bibnamefont {Huang}}, \ and\
  \bibinfo {author} {\bibfnamefont {P.~R.}\ \bibnamefont {Prucnal}},\
  }\href@noop {} {\bibfield  {journal} {\bibinfo  {journal} {Optics letters}\
  }\textbf {\bibinfo {volume} {45}},\ \bibinfo {pages} {4819} (\bibinfo {year}
  {2020})}\BibitemShut {NoStop}%
\bibitem [{\citenamefont {Borghi}\ \emph {et~al.}(2021)\citenamefont {Borghi},
  \citenamefont {Bazzanella}, \citenamefont {Mancinelli},\ and\ \citenamefont
  {Pavesi}}]{borghi-2021}%
  \BibitemOpen
  \bibfield  {author} {\bibinfo {author} {\bibfnamefont {M.}~\bibnamefont
  {Borghi}}, \bibinfo {author} {\bibfnamefont {D.}~\bibnamefont {Bazzanella}},
  \bibinfo {author} {\bibfnamefont {M.}~\bibnamefont {Mancinelli}}, \ and\
  \bibinfo {author} {\bibfnamefont {L.}~\bibnamefont {Pavesi}},\ }\href@noop {}
  {\bibfield  {journal} {\bibinfo  {journal} {Optics Express}\ }\textbf
  {\bibinfo {volume} {29}},\ \bibinfo {pages} {4363} (\bibinfo {year}
  {2021})}\BibitemShut {NoStop}%
\bibitem [{\citenamefont {Kim}\ and\ \citenamefont {Hong}(2024)}]{kim-2024}%
  \BibitemOpen
  \bibfield  {author} {\bibinfo {author} {\bibfnamefont {Y.}~\bibnamefont
  {Kim}}\ and\ \bibinfo {author} {\bibfnamefont {S.-H.}\ \bibnamefont {Hong}},\
  }\href@noop {} {\bibfield  {journal} {\bibinfo  {journal} {Nanophotonics}\ }
  (\bibinfo {year} {2024})}\BibitemShut {NoStop}%
\bibitem [{\citenamefont {Pruessner}\ \emph {et~al.}(2007)\citenamefont
  {Pruessner}, \citenamefont {Stievater},\ and\ \citenamefont
  {Rabinovich}}]{pruessner-2007}%
  \BibitemOpen
  \bibfield  {author} {\bibinfo {author} {\bibfnamefont {M.~W.}\ \bibnamefont
  {Pruessner}}, \bibinfo {author} {\bibfnamefont {T.~H.}\ \bibnamefont
  {Stievater}}, \ and\ \bibinfo {author} {\bibfnamefont {W.~S.}\ \bibnamefont
  {Rabinovich}},\ }\href@noop {} {\bibfield  {journal} {\bibinfo  {journal}
  {Optics letters}\ }\textbf {\bibinfo {volume} {32}},\ \bibinfo {pages} {533}
  (\bibinfo {year} {2007})}\BibitemShut {NoStop}%
\end{thebibliography}%

\end{document}